\documentclass[twocolumn,showpacs,preprintnumbers,amsmath,amssymb]{revtex4}

\usepackage{graphicx}
\usepackage{dcolumn}
\usepackage{bm}

\begin{document}

\preprint{APS/123-QED}

\title{Generation and detection of non-Abelian matrix Berry phases through manipulation of electric
confinement potential of a semiconductor quantum dot}

\author{S.-R. Eric Yang\footnote{ eyang@venus.korea.ac.kr}}
\author{N.Y. Hwang}
\affiliation{Physics Department, Korea  University, Seoul, Korea 
}


\begin{abstract}
A matrix Berry phase can be generated and detected by {\it all electric means} in II-VI or III-V n-type  semiconductor quantum dots by changing 
the shape of the confinement potential.  This follows from general symmetry considerations in the presence of spin-orbit coupling terms.
The resulting $2\times 2$ matrix Berry phase can be characterized by two  numbers of geometric origin.
We investigate how these parameters depend on 
the  shape and area of closed adiabatic paths.
We suggest how the matrix Berry phase may be detected in transport measurements.

\end{abstract}
\maketitle

\section{Introduction}

\begin{figure}[hbt]
\begin{center}
\includegraphics[width = 0.35\textwidth]{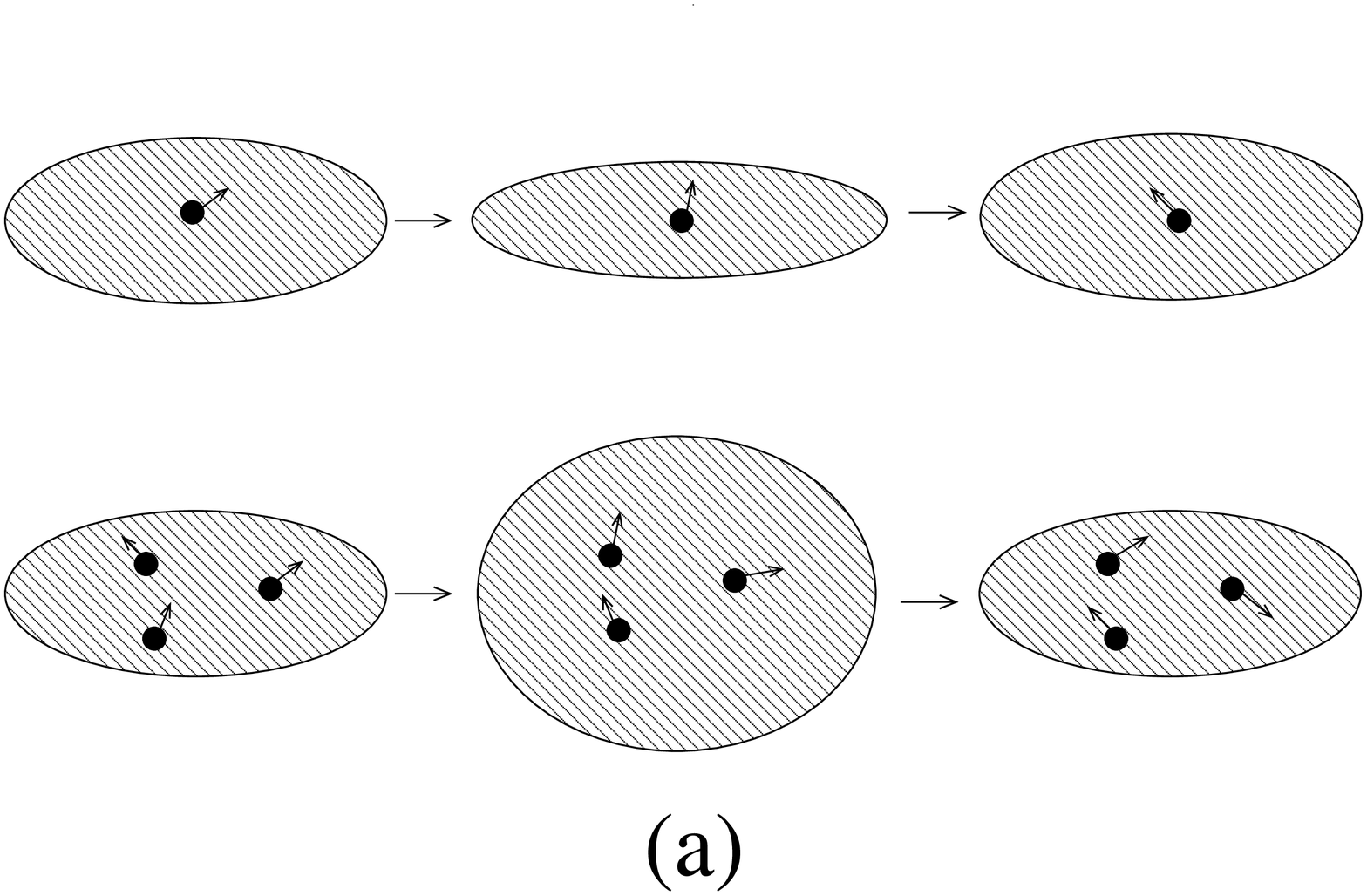}
\includegraphics[width = 0.2 \textwidth]{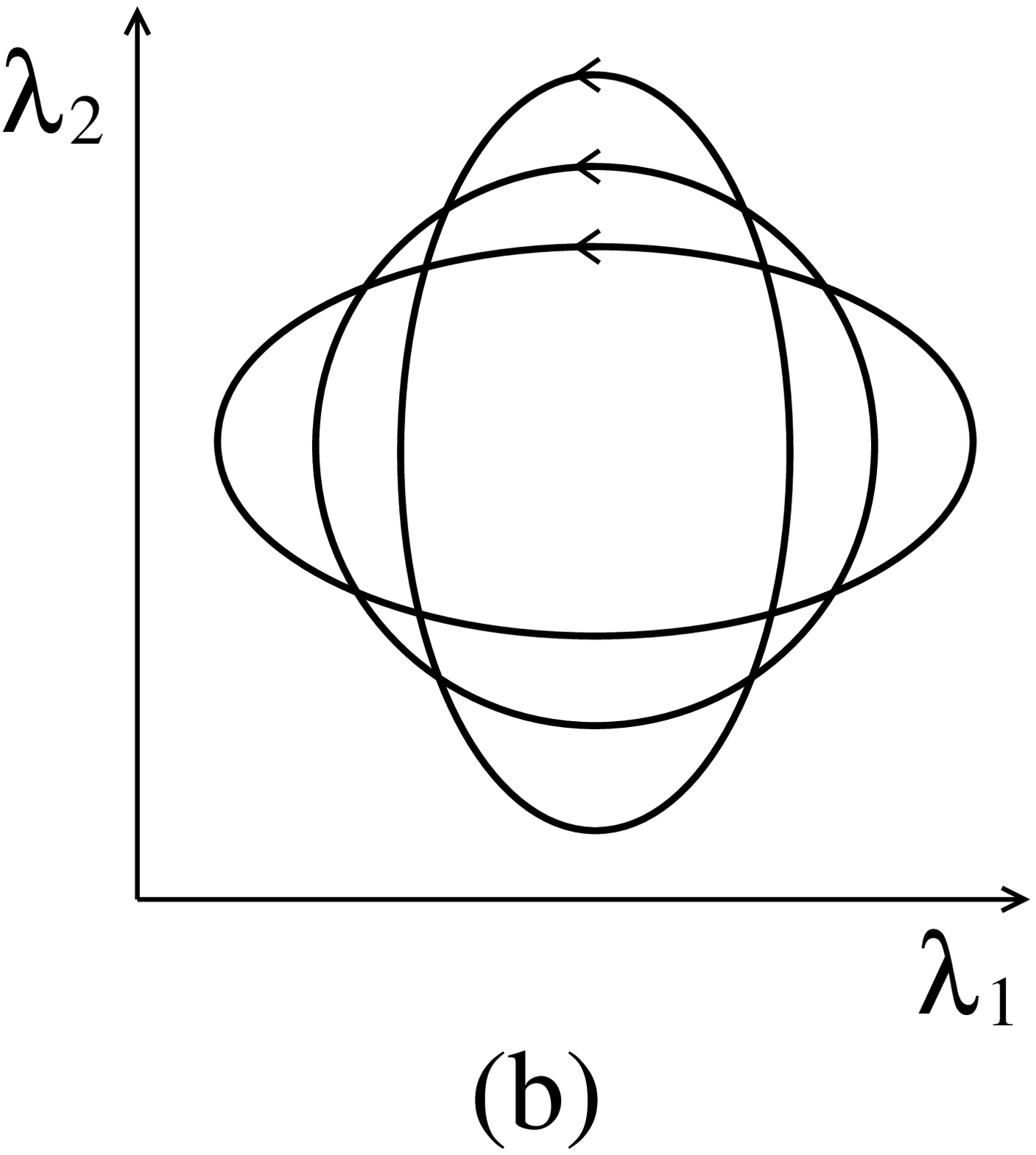}
\caption{
(a) Displays electrons in quantum dots.  A cyclic adiabatic change of the shape of the electric confinement potential  
does not return electron spins to the initial values.
(b) Several adiabatic paths are shown. Adiabatic parameters  $\lambda_1$ and  $\lambda_2$ provides a means to  control the shape of the
dot electrically.
}
\label{fig:shape1}
\end{center}
\end{figure}

\begin{figure}[hbt]
\begin{center}
\includegraphics[width = 0.35\textwidth]{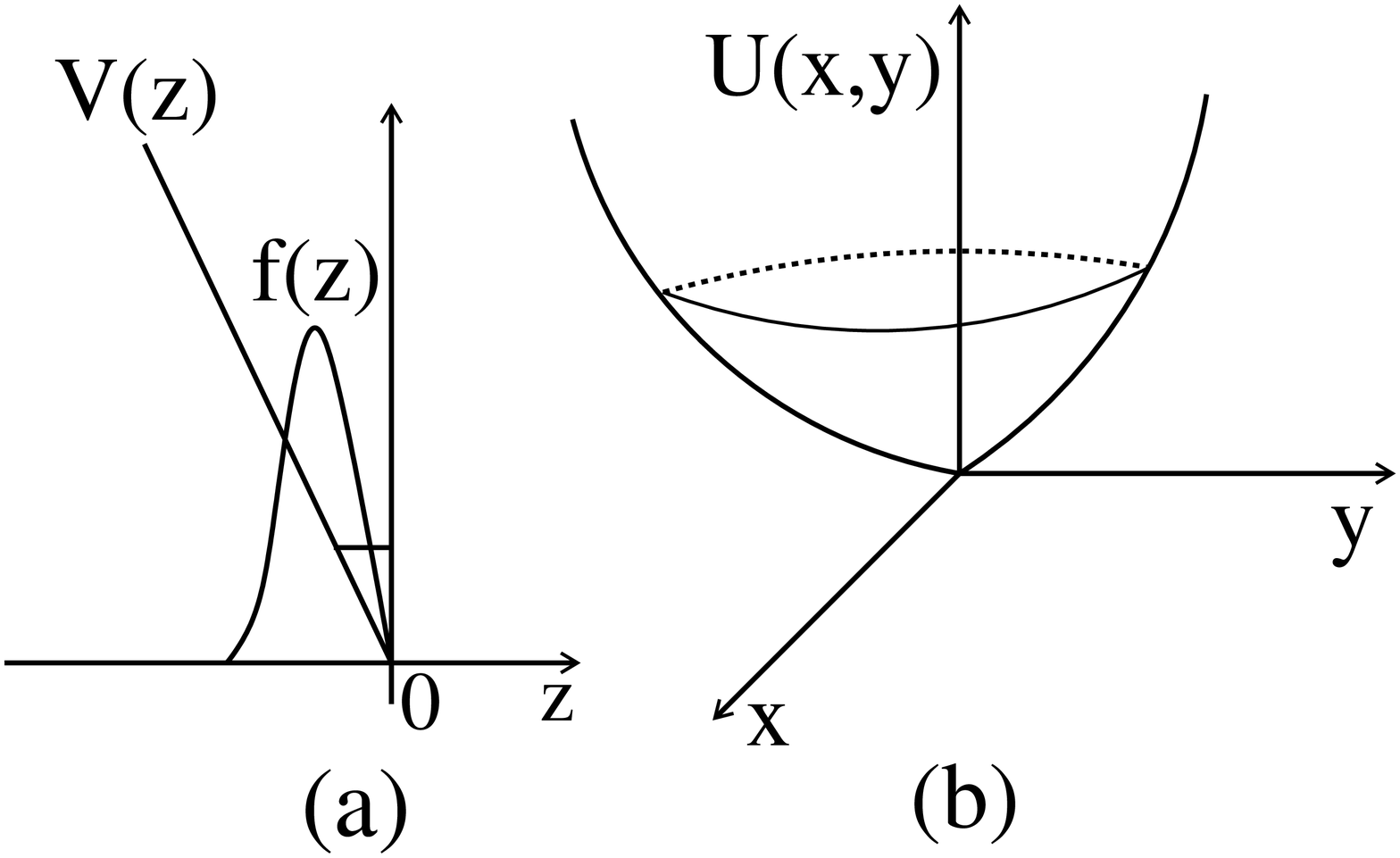}
\includegraphics[width = 0.15 \textwidth]{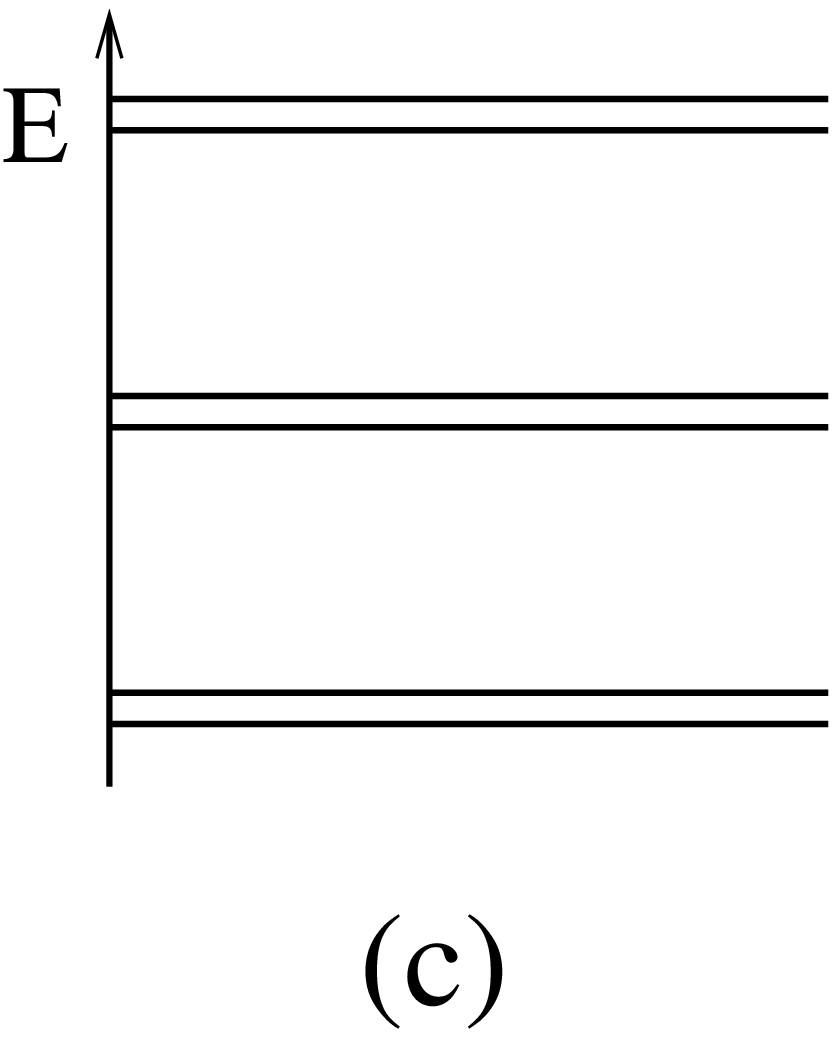}
\caption{Basic mechanism of adiabatic control is based on non-trivial degeneracy of II-VI and III-V semiconductors.
(a) An electric field along the z-axis quantizes the electronic motion in a triangular potential along the axis.
An adiabatic change can be induced by changing the magnitude of the electric field.
(b) The two-dimensional electronic motion is quantized in a distorted parabolic potential.
An adiabatic change can be induced by changing the magnitude of the distortion  potential.
(c)In the presence of the spin-orbit terms 
each discrete eigenstate  of a semiconductor quantum dot has
a double degeneracy due to time reversal symmetry in the absence of a magnetic field.} 
\label{fig:shape2}
\end{center}
\end{figure}

Single and few electron control by electric means in semiconductor quantum dots  would be valuable 
for spintronics, quantum information, and spin qubits\cite{Aws,Da,Loss}.
Adiabatic time evolution of degenerate eigenstates of a quantum system provides a means
for controlling individual quantum states through the generation of non-Abelian matrix Berry phases\cite{Wil,Sha,Bo}. 
This method can be used  to perform  universal quantum computation \cite{Zan}.
There are several semiconductor nanosystems  that exhibit matrix Berry phases: they include
excitons\cite{Sol},  CdSe nanocrystals\cite{Sere2}, and acceptor states of p-type semiconductors\cite{Bern}.
It is desirable to generate and detect matrix Berry phases by {\it all electric means} in semiconductor nanosystems.
Recently we have demonstrated 
theoretically that it is possible to control {\it electrically} electron spins of II-V and III-V
n-type semiconductor quantum dots\cite{yang1,yang2} and rings\cite{yang3} 
by changing the shape of the electric  confinement potential.  
The mechanism is based on  spin orbit couplings which generate  non-Abelian vector potentials.
The main ingredients of it are:
[a] Time-reversal symmetry, which leads to double degeneracy\cite{val,yang1},
which is depicted schematically in Fig.\ref{fig:shape1}.
[b] Non-Abelian U(2) gauge theory.
[c] Adiabatic changes that  break the parity symmetry of the electric confinement potential  of the dot.
When several electrons are present the matrix Berry phase can be generated only  when odd number of electrons are in the dot.
In such a case the effect of correlations  can be included in a compact way
(exchange effects do not affect the  non-Abelian vector potentials)\cite{yang2}.
It should be emphasized that  the presence of the matrix Berry phase 
follows from  general symmetry considerations.  The  matrix Berry phase can be present even in the absence of the spin-orbit terms.
A noteworthy example is semiconductor quantum
dot pumps\cite{Th,Br}, which  can be understood as a manifestation of a  matrix Berry phase\cite{yang4}.


Let us give a brief explanation of matrix Berry phases.
The electron state at time $t$ is given by
\begin{eqnarray}
|\Psi(t)\rangle =c_1(t)|\psi_1(t)\rangle +c_2(t)|\psi_2(t)\rangle,
\label{eq:instan}
\end{eqnarray}
where 
 $|\psi_{1,2}(t)\rangle$ are the  instantaneous basis states
satisfying $H(t)\psi_i(t)=E(t)\psi_i(t)$ for  $i=1,2$ ($H(t)$ is the Hamiltonian with the eigenenergy $E(t)$ at time $t$).
The matrix Berry phase $\Phi_C(1)$ connects $(c_1(T),c_2(T))$ to $(c_1(0),c_2(0))$, and is given by
\begin{eqnarray}
\Phi_C(1)
&=&e^{ i \sum_p A_p(t_n) d\lambda_p}....e^{ i  \sum_pA_p(t_1) d\lambda_p}\nonumber\\
&=&Pe^{ i \oint_C\sum_p A_p d\lambda_p},
\label{eq:matB}
\end{eqnarray}
where time slices are  ordered as $t_1<...<t_n<...$. The  
matrix vector potentials $A_p$ 
\begin{eqnarray}
(A_p)_{i,j}=i \langle\psi_i|\frac{\partial\psi_j}{\partial\lambda_p}\rangle,
\end{eqnarray}
are integrated along the path C in the parameter space in the order of increasing time. 
The path C is parameterized in time as $(\lambda_1(t),\lambda_2(t))$.
The matrix Berry phase is independent of the functions $\lambda_i(t)$ as long as they describe the same path C.

In applying matrix Berry phases to II-VI and III-V semiconductor quantum dots there are several issues that need to be addressed.
One issue is  whether the line integral along C 
can be {\it generally} converted into an areal integration over A in the expression for the matrix Berry phase.
For this purpose it is useful to consider 
the field strength 2-form  
\begin{eqnarray}
F=\sum_{\mu \nu} F_{\mu \nu} \mathrm{d}\lambda_\mu \wedge \mathrm{d}\lambda_\nu,
\end{eqnarray}
where
\begin{eqnarray}
F_{\mu \nu}=\partial_\mu A_\nu-\partial_\nu A_\mu +[A_\mu,A_\nu]
\end{eqnarray}
with $\partial_\mu A_\nu=\frac{\partial A_\nu}{\partial\lambda_\mu}$.
For a small area A it can be shown\cite{Sha} that the matrix Berry phase is 
\begin{eqnarray}
\Phi_C(1)=\exp(i A F_{12}).
\end{eqnarray}
For an arbitrarily  large A there are two cases:[1]
If $[A_\mu,A_\nu]=0$, the curvature reduces to $F_{\mu \nu}=\partial_\mu A_\nu-\partial_\nu A_\mu $
and we can use Stokes' theorem to convert the line integral along C to an areal integration over A
\begin{eqnarray}
\Phi_C(1)=\exp(i\frac{1}{2}\int_A F_{\mu\nu}\mathrm{d}\lambda_\mu \wedge \mathrm{d}\lambda_\nu).
\label{eq:stokes}
\end{eqnarray}
[2] When $[A_\mu,A_\nu]\neq0$  it is not possible to convert the line integral into an areal integration.
We will investigate the applicability of Stokes' theorem in the case of Rashba and Dresselhaus terms.

Second issue is that it is often complicated to calculate the non-Abelian vector potentials,
and, consequently, 
it  is difficult to compare    experimental and theoretical results.
This is  especially so 
when many electrons are present. 
It would be useful to have a simple, yet  general, mathematical expression  for  the  matrix Berry phase. 
We will show that the matrix Berry phase is a unitary $2\times2$ matrix and that it may be thought of as a spin rotation matrix\cite{Ch}.
A spin rotation matrix is characterized by the direction and angle of rotation.
It is unclear how the geometric  properties of the adiabatic path  are reflected
on them.  We show in this paper how  the information of the path C can be encoded in the direction and angle of the rotation.

Another issue is whether 
the matrix Berry phase can be  detected
in  a  transport measurement, which would complement the  detection of
matrix Berry phases  by infrared optical measurements\cite{yang1}.

\begin{figure}[hbt]
\begin{center}
\includegraphics[width = 0.4\textwidth]{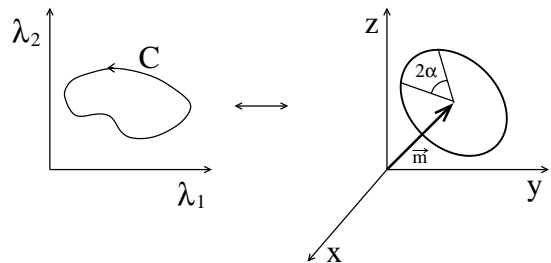}
\caption{The geometric information about the closed curve in the parameter space
is encoded into the rotation axis $\vec{m}$ and angle $2\alpha$.
When only the Rashba term is present $\vec{m}$ is independent of path C.}
\label{fig:shape3}
\end{center}
\end{figure}

\begin{figure}[hbt]
\begin{center}
\includegraphics[width = 0.35 \textwidth]{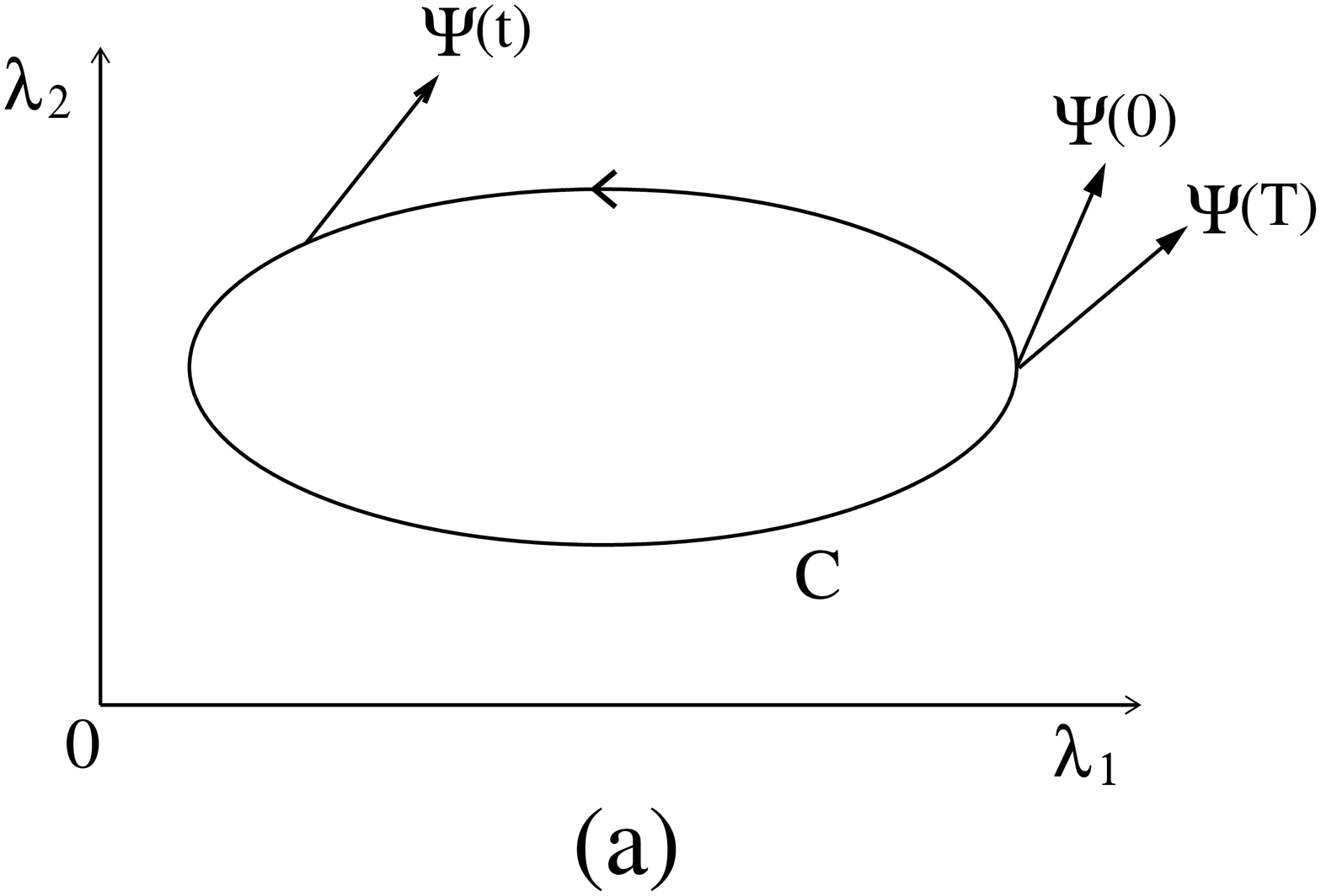}
\includegraphics[width = 0.25 \textwidth]{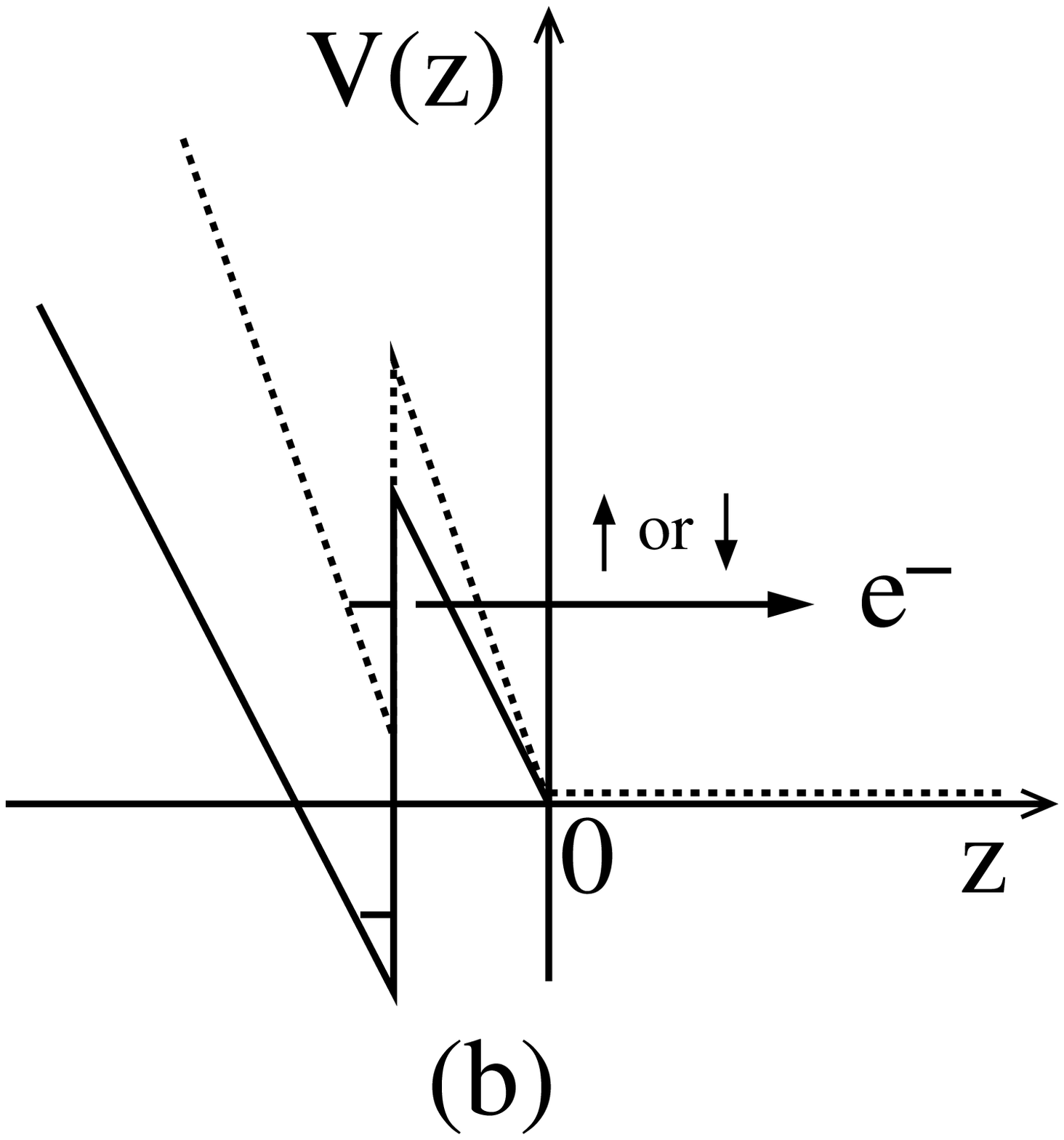}
\caption{(a) A matrix Berry phase is generated when  the adiabatic parameters $\lambda_1$ and
$\lambda_2$ go through a cyclic change.
(b)After a matrix Berry phase is generated the Rashba electric field is suddenly increased
so that an electron can tunnel  out the dot. }
\label{fig:potential_shape}
\end{center}
\end{figure}

Let us now give a brief summary of the main results of this paper.
First, 
the matrix Berry phase has the form
\begin{eqnarray}
\Phi_C(n)=e^{\frac{i}{2}(2\alpha n)\vec m \cdot\vec \sigma} =\cos(\alpha n)I +i\sin(\alpha n)\vec m \cdot \vec \sigma,
\label{eq:matrix.Berry0}
\end{eqnarray}
where $\vec \sigma$ are  Pauli spin matrices and  $n$ is the number of periodic adiabatic cycles.
The information about the {\it closed} curve in the parameter space
is  encoded into the rotation axis 
\begin{eqnarray}
\vec m=(\text{Re}(\beta),-\text{Im}(\beta),\sqrt{1-|\beta|^2})
\end{eqnarray}
and the angle of rotation  
$2\alpha n$. 
The constants $\alpha $ and $\beta$ are
real and  complex numbers, respectively.
This result  is even valid  for  many electron systems with  correlations as long as double degeneracy is present.
The expression Eq.(\ref{eq:matrix.Berry0})  is valid {\it only} at  time  
$t=nT$, and it 
should be stressed that  in the time interval $(n-1)T<t<nT$ the probability amplitudes 
$c_1(t)$ and $c_2(t)$ exhibit a much more complicated
behavior, see Fig.\ref{fig:c1square.c1square.fit.Dress}.  
Second,
we find $\beta=1$ and $\vec m=(1,0,0)$ when only the Rashba term is present, and all  the geometric 
information is contained in the rotation angle $\alpha$.
We have derived  an  analytical expression for $\alpha$.  Furthermore,  
remarkably $[A_\mu,A_\nu]=0$ and Stokes' theorem can be applied.
When both Rashba and Dresselhaus terms are present each path has   different  $\vec m$ and $\alpha $.
We have investigated how they depend on the shape of each path.
Moreover,  we find  that $[A_\mu,A_\nu]\neq0$ and  that Stokes' theorem cannot be applied.
Third, we propose how the presence of the matrix Berry phase can be measured in transport experiments by 
changing the electric confinement potential suddenly, see Sec.III.  This is shown schematically
in Fig.\ref{fig:potential_shape}.

\section{Semiconductor quantum dots with the Rashba and Dresselhaus spin orbit couplings}

\subsection{Model}

The Hamiltonian is 
$H=-\frac{\hbar^2\nabla^2}{2m^*}+U(x,y)+V(z)+H_\mathrm{R}+H_\mathrm{D}$.
We take the two-dimensional potential to be $U(x,y)=\frac{1}{2}m^*\omega^2_x x^2+\frac{1}{2}m^*\omega^2_y y^2+
\epsilon' y$, where the term $\epsilon' y$ represents a distortion of the two-dimensional harmonic potential.
The strengths of the harmonic potentials are denoted by $\omega_x$ and $\omega_y $.   
The characteristic lengths scales along x- and y-axis are $R_{x,y}=\sqrt{\frac{\hbar}{m^*\omega_{x,y}}}$.
In our work  the triangular potential $V(z)$ is sufficiently strong and  only the lowest energy subband is included.
The characteristic length scale along the z-axis is $1/\sqrt{0.8(2meE/\hbar^2)^{2/3}}$,
where $E$ is the Rashba electric field applied along the z-axis.
The Rashba spin orbit term\cite{ras,Nitta} is 
\begin{eqnarray}
H_\mathrm{R}=c_\mathrm{R} \left( \sigma_x k_y -\sigma_y k_x \right),
\end{eqnarray}
and the Dresselhaus spin orbit term\cite{dre} is
\begin{eqnarray}
H_\mathrm{D}=c_\mathrm{D}\left( 
\left( \sigma_x k_x \left(k_y^2-k_z^2 \right) \right)
+\left( \sigma_y k_y \left(k_z^2-k_x^2 \right) \right)
\right).
\label{Dresselhaus}
\end{eqnarray}
Note that the  Rashba spin orbit constant $c_R$ {\it depends on the external electric field} $E$.
Here  $k_{x,y}$ are momentum operators 
($k_x=\frac{1}{i}\frac{d}{dx}$
and similarly with $k_y$.).  
The Hamiltonian matrix is represented
in the basis states $|m n \sigma \rangle $ of 
the harmonic oscillator states in the xy plane with
spin component $\sigma $.  The subband wavefunction $f(z)$ is suppressed in the notation $|m n \sigma \rangle$.
Let us take $\omega_x=2\omega_y$ (Other values of $\omega_x$ can also be chosen). Then the lowest eigenenergy state
of $H$ is $|mn\rangle=|00\rangle$ with the energy $E_0=\frac{3}{2}\hbar\omega_y$ and
the next lowest eigenenergy state  is $|01\rangle$ with the energy $E_1=\frac{5}{3}E_0$.
The typical value of the energy spacing between the quantum dot levels, $E_0$, is of  order $1-10 meV$.
It can be several times larger in self-assembled dots.
The energy scale of the Rashba term is $E_R=c_R/ R\sim 0.01-10meV$, where the   length
scale $R\sim 100 \AA$ is the lateral dimension of the quantum dot.
The  energy scale of the distortion potential  
is $E_p=\langle 0|\epsilon' y|1\rangle$.  Its 
magnitude is of order $1-10meV$, depending on the electric field applied along the y-axis.
The energy scale of the Dresselhaus term is $E_D=c_D/ R^3$, and
it can be larger or smaller than the Rashba term, depending on the material\cite{ras}.
Here $E_D=ic_D\langle 0|k_y|1\rangle(\langle f(z)|k_z^2|f(z)\rangle-\langle 0|k_x^2|0\rangle)=c_D/R_yR_z^2-c_D/R_yR_x^2$, and it originates from the 
second term of the Dresselhaus term (the first term in the Dresselhaus term is zero in our model since $\langle 0|k_x|0\rangle=0$).

We will use a  truncated version of the Hamiltonian matrix, which makes it possible 
to  write  the eigenstates   as a  linear combination
of four basis states made out of $|mn \rangle$ 
and spin degree of freedom:
\begin{eqnarray}
| \psi \rangle=
&&c_{0,0,\uparrow}|0,0, \uparrow \rangle+c_{0,1,\uparrow}|0,1, \uparrow \rangle \nonumber\\ 
&+&c_{0,0,\downarrow}|0,0, \downarrow \rangle+c_{0,1,\downarrow}|0,1, \downarrow \rangle.
\end{eqnarray}
The advantage of this truncated model is that it is {\it exactly solvable}.
In this paper we choose the following  eigenstates of the lowest energy shell as the  instantaneous  basis vectors: 
\begin{eqnarray}
|\psi_1\rangle&=&
\frac{1}{N_1}
\left(
\begin{array}{c}
3 E_p \\
E_0-\sqrt{E_0^2+9(E_D^2+E_p^2+E_R^2)} \\
3(E_D-i E_R) \\
0
\end{array}
\right),\nonumber\\
|\psi_2\rangle&=&|\overline{\psi}_1\rangle.
\label{eq:lin_comb2}
\end{eqnarray}
The non-Abelian vector potentials are evaluated with respect to these basis vectors.

In the following we choose the adiabatic parameters
as the Rashba energy and  the distortion energy
\begin{eqnarray}
\lambda_1&=&E_R=c_R/(\sqrt{2}R_y)\nonumber\\
\lambda_2&=&E_p=\langle 0|\epsilon' y|1 \rangle=\frac{\epsilon'}{\sqrt 2} R_y
\end{eqnarray}
The cyclic adiabatic path is given by
$E_R(t)=E_{R,c}+\Delta E_R \cos(\omega t)$ and   $E_p(t)= E_{p,c}+\Delta E_p \sin(\omega t)$.

\subsection{Geometric dependence}

\ \\
\ \\
\begin{figure}[hbt]
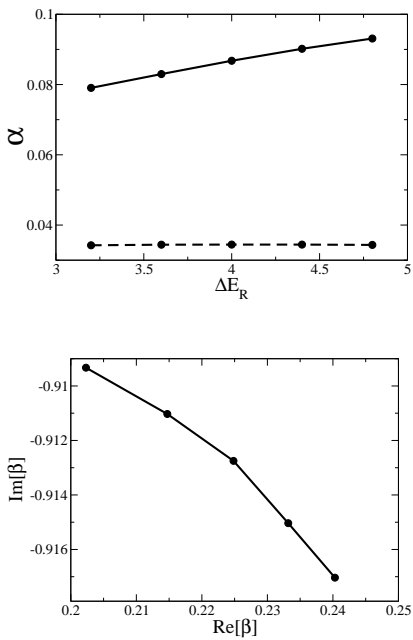

\begin{center}
\includegraphics[width = 0.30 \textwidth]{alpha.ds.and.or.rs.eps}
\ \\
\ \\
\ \\
\includegraphics[width = 0.30 \textwidth]{beta2.eps}
\caption{For elliptic paths with fixed area with $\Delta E_R$ changing  from $3.2 E_0$ to $4.8 E_0$.
In the presence of both of the Dresselhaus and Rashba terms (solid line).
Only the Rashba term present (dashed).
How $\beta$ changes with semiaxis $\Delta E_R$ in the presence of both of the Dresselhaus and Rashba terms.}
\label{fig:geometry}
\end{center}
\end{figure}

When only the Rashba term is present the structure of the non-Abelian vector potentials is remarkably simple.
The orthonormalization  $\langle\psi_i|\psi_j\rangle=\delta_{ij}$    gives that the diagonal matrix elements
$(A_p)_{i,i}$ are real   and that the off-diagonal elements satisfy $(A_p)_{i,j}=(A_p)^*_{j,i}$.
We find non-Abelian vector potentials $A_1$ and $A_2$
\begin{eqnarray}
&&A_1=-\frac{E_p \left(1+\frac{1}{\sqrt{9 E_p^2+9 E_R^2+1}}\right)}{2 \left(E_p^2+E_R^2\right)} \sigma_x,\\
&&A_2=\frac{E_R \left(1+\frac{1}{\sqrt{9 E_p^2+9 E_R^2+1}}\right)}{2 \left(E_p^2+E_R^2\right)} \sigma_x.
\end{eqnarray}
(Note that $\langle f(z)|\frac{\partial}{\partial E_R}|f(z)\rangle =0$.)
The diagonal elements $A_{ii}$ are  zero and the off-diagonal elements $A_{ij}$ are real.
From this it follows that the parameter $\beta=1$ and only one geometric number $\alpha $ is needed.
The matrix Berry phase is given by
\begin{eqnarray}
\Phi_C(n)
=
\left(
\begin{array}{cc}
\cos(\alpha n) & i\sin(\alpha n) \\
i\sin(\alpha n)& \cos(\alpha n)
\end{array}
\right). 
\label{eq:matrix.Berry7}
\end{eqnarray}
One can show that $[A_1,A_2]=0$ when
only the Rashba term is present, and that 
Stokes' theorem can be applied,  Eq. (\ref{eq:stokes}).
The field strength has a  simple form 
$F_{12}=f\sigma_x/E_0^2$ where
\begin{eqnarray}
f=-\frac{9}{2 \left(9 x^2+9 y^2+1\right)^{3/2}}, 
\end{eqnarray}
with
$x=E_p/E_0$  and $y=E_R/E_0$.
From this we can compute the rotation angle    
\begin{eqnarray}
\alpha=\int_{A}f d x d y,
\label{eq:integ}
\end{eqnarray}
which gives $|\alpha|<0.7854$.

When both the Rashba and Dresselhaus terms are present Stokes' theorem is not applicable.
We have  
investigated how $\vec{m}$ and $\alpha$ depend on the shape of the paths.
They can be  computed numerically by solving the time dependent Schr{\"o}dinger equation
\begin{eqnarray}
i\hbar\dot{c}_i=-\sum_j A_{ij}c_j.
\label{Sch}
\end{eqnarray}
The matrix elements $A_{ij}$ are given by
$A_{ij}= \hbar \sum_p(A_p)_{i,j}\frac{d\lambda_p}{dt}$, where the
sum over $p$ in $A_{ij}$ is meant to be the sum  over $\lambda_p$.
We have considered elliptic paths with the
area $\pi \Delta E_R \Delta E_p=16 \pi E_0^2$ with $E_{R,c}=5E_0$  and $E_{p,c}=5E_0$.
The results are shown in Fig.\ref{fig:geometry}.
(Note that the dependence of $E_D$ on the adiabatic parameter $E_R$ is not well-known, but
the essential physics does not depend on the exact functional form, as discussed in  Ref.\cite{yang1}).
Fig.\ref{fig:geometry} also shows numerical results for $\alpha$ when only the Rashba term is present. 
We have {\it verified}
that these numerical results agree well
with the results obtained from the equation Eq.(\ref{eq:integ}).

\subsection{Time dependence  }

\ \\
\ \\
\begin{figure}[hbt]
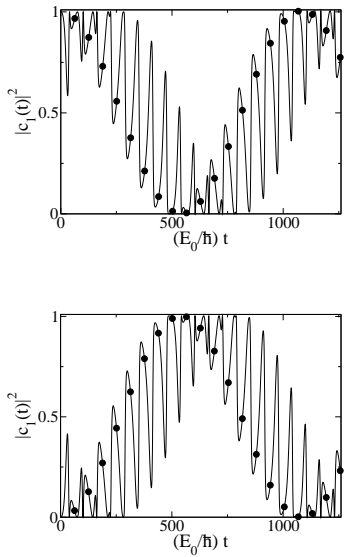

\begin{center}
\includegraphics[width = 0.25 \textwidth]{c1.square.fitb.eps}
\ \\
\ \\
\ \\
\includegraphics[width = 0.25 \textwidth]{c2.square.fitb.eps}
\ \\
\caption{ Probabilities as a function of time in the presence of  the Rashba spin orbit coupling.
Black dots are fitted values at $t=nT$ using  the analytical expression for the matrix Berry phase, Eq.(\ref{eq:matrix.Berry7}).}
\label{fig:c1square.c2square.fit}
\end{center}
\end{figure}

\ \\
\begin{figure}[hbt]
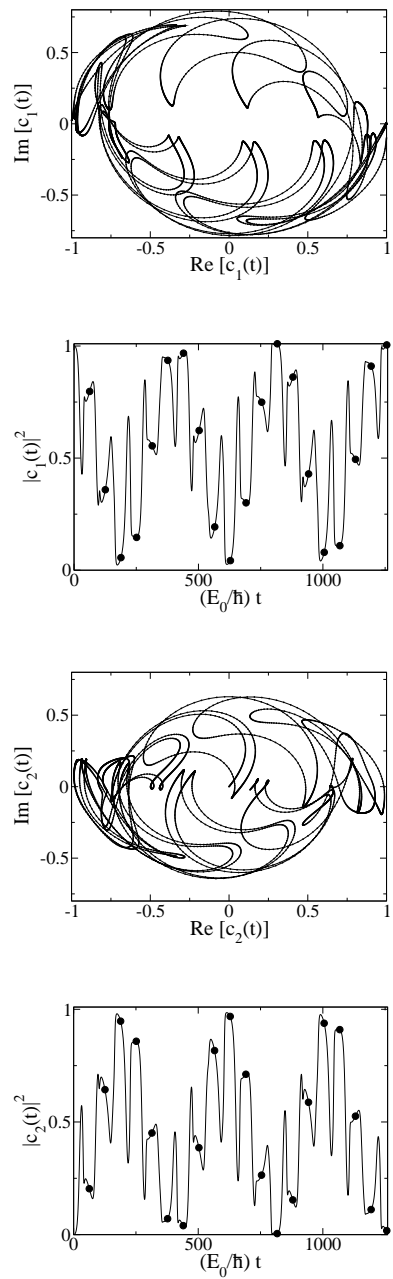

\begin{center}
\ \\
\includegraphics[width = 0.28 \textwidth]{c1comp.eps}
\ \\
\ \\
\ \\
\includegraphics[width = 0.28 \textwidth]{c1.square.Dress.fitb.eps}
\ \\
\ \\
\ \\
\includegraphics[width = 0.28 \textwidth]{c2comp.eps}
\ \\
\ \\
\ \\
\includegraphics[width = 0.28 \textwidth]{c2.square.Dress.fitb.eps}
\ \\
\caption{Probability $c_1(t)$ when Dresselhaus spin orbit interaction is included.
Black dots are fitted values using  the analytical expression for the matrix Berry phase,  Eq.(\ref{eq:matrix.Berry0}).
The initial state is   $c_1(0)=1$ and  $c_2(0)=0$.}
\label{fig:c1square.c1square.fit.Dress}
\end{center}
\end{figure}

Let us first investigate the time evolution of  the system  in the presence of the Rashba term.
The time-dependent Schr{\"o}dinger equation is solved using  the Runge-Kutta method.
For the parameters $E_{R,c}= 2 E_0$, $E_{p,c}=E_0$, $\Delta E_R= 1.9 E_0$, $\Delta E_p= 0.9 E_0 $, 
and $\hbar\omega_1=0.1 E_0$ we calculate  $|c_1(t)|^2$ and $|c_2(t)|^2$, see  Fig.\ref{fig:c1square.c2square.fit}.
We can fit the data to $|c_1(n T)|^2=\cos^2(\alpha n)$,
$|c_2(n T)|^2=\sin^2(\alpha n)$, 
with $\alpha \approx 0.182$.
In this case the matrix Berry phase takes a simple form with $\beta=1$.
We have {\it tested} our numerical method as follows: For two different periods $T_1$ and $T_2$
we  find the matrix Berry phases are the same, i.e., $c_2(T_1)=c_2(T_2)$  and   $c_1(T_1)=c_1(T_2)$.
This is consistent with the fact that the matrix Berry phase is a geometric effect independent
of how the path is parameterized.
Let us now include the Dresselhaus spin orbit term as described in Ref \cite{yang1}.
In this case the matrix Berry phase  is more complicated since $\beta\neq 1$.
The diagonal elements $A_{ii}$ are  non-zero and the off-diagonal elements $A_{ij}$ are complex numbers.  
Using the same parameters $E_{R,c}$, $E_{p,c}$, $\Delta E_R$, $\Delta E_p$, and $\omega$ 
as in the previous calculation, we have computed  $|c_1(t)|^2$ and $|c_2(t)|^2$,
which are displayed in Fig.\ref{fig:c1square.c1square.fit.Dress}.
These results can be fitted with
the parameters $\alpha \approx -0.477$ and $\beta \approx -0.190 +i 0.965$.
In both cases the numerical results are consistent with those of  the analytical expression
at  $t=nT$, given by Eq.(\ref{eq:matrix.Berry0}).  In the time interval $(n-1)T<t<nT$ the coefficients $c_{1,2}(t)$ display a complicated behavior. 
It should be stressed that even when the adiabatic path is {\it not} closed the actual state $\Psi(t)$ is given by a nontrivial 
linear combination of the instantaneous eigenstates $\psi_1(t)$ and $\psi_2(t)$.  In this case a  matrix phase of the form 
\begin{eqnarray}
\Phi_C=
Pe^{ i \int_C\sum_p A_p d\lambda_p},
\end{eqnarray}
will appear, where the path $C$ is unclosed.

\section{Discussions and conclusions}

We have investigated II-VI and III-V n-type semiconductor dots with spin orbit coupling terms
and  have shown in this paper that the geometric
information of a closed adiabatic path can be encoded in the rotation axis and angle of the spin half matrix.
This result  is applicable even for strongly correlated states, as along as the states are doubly degenerate.
In our truncated model for the Hamiltonian the matrix Berry phase can be found {\it exactly} when  the Rashba term is present.
In this case the Stokes' theorem is applicable and the relevant field strength can be calculated analytically.

Even when the adiabatic path is {\it not} closed the  actual state,$\Psi(t)$, at time $t$ will not be the instantaneous eigenstate
$\psi_i(t)$.
Instead it is  a  linear combination of the instantaneous eigenstates
\begin{eqnarray}
|\Psi(t)\rangle =c_1(t)|\psi_1(t)\rangle +c_2(t)|\psi_2(t)\rangle.
\label{eq:instan}
\end{eqnarray}
The expansion coefficients $c_i(t)$ change in a non-trivial way in the interval  $(n-1)T<t<nT$ ,
see Figs.\ref{fig:c1square.c2square.fit} and \ref{fig:c1square.c1square.fit.Dress}. 

Experimental investigations of matrix Berry phases in II-VI and III-V  semiconductor quantum dots would be most interesting.
Optical dipole measurements\cite{hei} were proposed before\cite{yang1}.   Here we propose a transport measurement.
In order to measure the matrix Berry phase we  perform the following set of  manipulations:
\begin{enumerate}
\item  Prepare an initial state by applying a small magnetic field and taking the zero field limit, as explained in detail in Ref. \cite{yang1}.

\item Perform an adiabatic cycle by changing $E_R$ and $E_p$ that defines the shape of the electric confinement potential,
see Fig.\ref{fig:potential_shape}(a).

\item Change suddenly the  electric field along the z-axis so that the electron can tunnel out of the dot, see Fig.\ref{fig:potential_shape}(b).

\item Measure the  spin dependence of the  tunneling current\cite{Da}.

\item Repeat this set of procedures many times.
\end{enumerate}
After this procedure we repeat the entire scheme one more time except the item 2.
The difference between spin-up (down) currents of these two schemes is a demonstration of the presence of  the matrix Berry phase.

Our truncated model can be applied to  self-assembled quantum dots\cite{pet}, which have  large energy shell differences.
In these dots exchange and correlation effects can be neglected\cite{yang2},
and it would be sufficient to consider only the electron in the last occupied  energy shell.
On the other hand, in  gated quantum dots\cite{Ka} with smaller energy differences between the shells and with odd number of electrons 
correlation effects may be relevant.
A computational scheme to calculate the matrix Berry phase in such a system is developed in Ref.\cite{yang2}.

\begin{acknowledgments}
This work was  supported by grant No. R01-2005-000-10352-0 from the Basic Research Program
of the Korea Science and Engineering
Foundation and by Quantum Functional Semiconductor Research Center (QSRC) at Dongguk University
of the Korea Science and Engineering
Foundation. In addition this work was supported by The Second Brain Korea 21 Project. 
\end{acknowledgments}


\begin{references}
\bibitem{Aws}D. D. Awschalom, D. Loss, and N. Samarth, Semiconductor Spintronics and Quantum Computation (Springer, Berlin, 2002).
\bibitem{Da}S. Datta and B. Das, Appl Phys. Lett.  {\bf 56}, 665 (1990).
\bibitem{Loss}D.Loss and D.P. DiVincenzo, Phys. Rev. A {\bf 57}, 120 (1998); 
S.D. Lee, S.J. Kim, Y.B. Cho, J.B. Choi, Sooa Park, S.-R. Eric Yang, S.J. Lee, 
and T.H. Zyung, Appl Phys. Lett.  {\bf 89}, 023111 (2006).
\bibitem{Wil}F. Wilczek and A Zee, Phys. Rev. Lett. {\bf 52}, 2111 (1984).
\bibitem{Sha}{\it Geometric Phases in Physics}, edited by A. Shapere and F. Wilczek
(World Scientific, Singapore, 1989). 
\bibitem{Bo}{\it The goemetric phase in quantum systems}, A. Bohm, A. Mostafazadeh, H. Koizumi, Q.Niu, and J. Zwanziger (Springer-Verlag, Berlin 2003).
\bibitem{Zan}P. Zanardi and M. Rasetti, Phys. Lett. {\bf 264}, 94 (1999).
\bibitem{Sol} P. Solinas,  P. Zanardi,   N. Zanghi, F. Rossi, Phys. Rev.A {\bf 67}, 062315 (2003).
\bibitem{Sere2}Yu. A. Serebrennikov, Phys. Rev. B {\bf 70}, 064422 (2004). 
\bibitem{Bern}B. A. Bernevig and S.-C. Zhang, Phys. Rev. B {\bf 71}, 035303 (2005).
\bibitem{yang1}S.-R. Eric Yang and N.Y. Hwang,  Phys. Rev. B  {\bf 73}, 125330 (2006).
\bibitem{yang2}S.-R. Eric Yang,  Phys. Rev. B   {\bf 75}, 245328 (2007).
\bibitem{yang3}S.-R. Eric Yang,  Phys. Rev. B   {\bf 74}, 075315 (2006).
\bibitem{val}M. Valin-Rodriquez, A. Puente, and L. Serra, Phys. Rev. B   {\bf 69}, 085306  (2004).
\bibitem{yang4} N.Y. Hwang, S.C. Kim, P.S. Park, and S.-R. Eric Yang, arXiv:0706.0947.
\bibitem{Th} D.J. Thouless, Phys. Rev. B {\bf 27}, 6083 (1983).
\bibitem{Br} P. W. Brouwer, Phys. Rev. B {\bf 58}, R10135 (1998).
\bibitem{Ch} M.A. Nielson and I.L. Chung, Quantum Computation and Quantum Information, Cambridge, University Press (2000).
\bibitem{ras}E.I. Rashba, Physica E,  {\bf 34}, 31 (2006).
\bibitem{Nitta}J. Nitta, T. Akazaki, H. Takayana, and T. Enoki, Phys. Rev. Lett. {\bf 78}, 1335 (1997).
\bibitem{dre}G. Dresselhaus, Phys. Rev. {\bf 100}, 580 (1955).
\bibitem{hei}D. Heitmann and J. P. Kotthaus, Phys. Today  {\bf 56} (6), 56 (1993).
\bibitem{pet}P. M. Petroff, A. Lorke, and A. Imamoglu, Phys. Today  {\bf 54} (5), 46 (2001).
\bibitem{Ka} M.A. Kastner, Rev. Mod. Phys.  {\bf 64}, 849 (1992).

\end{references}
\end{document}